\documentstyle[prb,aps]{revtex}

\tighten

\begin{document}

%\draft

%\preprint{\sf Draft}
\title{A phenomenological electronic stopping power model
for molecular dynamics and 
Monte Carlo simulation
of ion implantation into silicon}

\author{David Cai$^{1}$, Niels Gr{\o}nbech-Jensen$^{1}$, 
Charles M. Snell$^{2}$, and
Keith M. Beardmore$^{1}$}

\address{$^{1}$ Theoretical Division, and 
 $^{2}$Applied Theoretical and Computational Physics Division, \\
Los Alamos National Laboratory, 
Los Alamos, New Mexico 87545}

\maketitle
\begin{abstract}
It is crucial to have a good phenomenological model of electronic
stopping power for modeling the physics of ion implantation into
crystalline silicon. In the spirit of the Brandt-Kitagawa effective
charge theory,
we develop a model for electronic stopping power for an ion, which
can be factorized into (i)
 a globally averaged effective charge taking into account
effects of close
and distant collisions by target electrons with the ion, and (ii)
a local charge density dependent electronic stopping power for a proton.
This phenomenological  model is implemented
into both molecular dynamics and Monte Carlo simulations. 
There is only
one free parameter in the model, 
namely, the one electron radius $r_{s}^{\circ}$ for unbound
electrons. 
By fine tuning
this parameter, it is shown that the model can work
successfully for both boron and arsenic implants.
We report that the results
of the dopant profile simulation for both species
are in excellent agreement with the
experimental profiles  measured by secondary-ion mass spectrometry
(SIMS)
over a wide range of energies and with different incident directions.
We point out that the model has wide applicability, for it captures
the correct physics of electronic stopping in
ion implantation. This model also provides
a good physically-based damping mechanism for molecular dynamics
simulations in the electronic stopping power regime,
as evidenced by the striking agreement of
dopant profiles calculated in
our molecular dynamics simulations with the SIMS data.

\end{abstract}


%\begin{multicols}{2}
%\narrowtext

\section{Introduction}
Ion implantation in semiconductors is an important
technology in integrated circuit device fabrication \cite{Ziegler-ST}.
 A reliable 
description of as-implanted profiles and the resulting
damage is needed for technological
development, such as device design and modeling, as well as
process optimization and control in the fabrication environment.
For semiconductor devices whose physical dimensions are of
order of submicrons or smaller,  low implant energies
and reduction of thermal processing  are necessary, resulting in
more prominent channeling effects in the
as-implanted profiles and less post-implant diffusion.
At these physical dimensions, it is essential to obtain
the two- or three-dimensional
details of the ever shallower and more compact dopant
and damage profiles  for post-implant diffusion
simulations.

Study of the energy loss of channeled
particles has a long history \cite{Robinson}, for the channeling
features can be used to elucidate the energy-loss mechanisms.
Earlier analytical treatments of the implant profiles based on moment
distributions, derived from the Lindhard-Scharff-Schiott theory (LSS)
\cite{LSS}, preclude  channeling because of the amorphous nature
of the targets assumed in the studies. Later,
it was realized  that, because of the
channeling effect, electronic stopping power plays a much more
significant role in ion implantation into crystalline solids
than otherwise would be deduced from the application \cite{LS}
of the LSS theory to
amorphous materials. It is  especially true
for heavy ion implants at low
energies, such
as arsenic ions in the energy range below $700$keV \cite{Yang-thesis,Sze}.
For implantation into silicon, most 
Monte Carlo (MC) models are only concerned with boron implants,
and have not modeled arsenic implants accurately with an electronic
stopping power model  consistent  with that used for boron
\cite{Yang,Murthy}. As will be shown below, the phenomenological model we
developed for electronic stopping power can be implemented into
a Monte Carlo simulation program for both boron and arsenic implants in
different channels
with equal success over a wide range of implant energies. 

In addition
to Monte Carlo simulations with the binary collision approximation (BCA)
\cite{RT},
 molecular dynamics (MD)
incorporating
multiple-interactions via  many-body potentials can also be used to
simulate the behavior of energetic ions in amorphous or crystalline
silicon. This method is especially applicable at low energies, for which
many-body, and multiple interactions are increasingly important \cite{Eckstein}. 
 Although it is well known that the BCA is valid for
 high incident energies ($\sim 0.1$keV up to $\sim$MeV, the upper limit is
set
by relativistic effects), in a cascade, especially initiated by
a relatively low
energy ion, the energy of the ions
 will decrease and eventually 
reach the lower validity limit of the BCA at which many-body effects become
important \cite{RT,Eckstein}. For crsytals of high symmetry, the BCA can be
modified to account for simultaneous collisions in 
channels\ \cite{RT,Robinson-Rad}, and MD results can provide good insight
into how to successfully modify the BCA in this situation.
Moreover, MD results can be compared to BCA Monte Carlo simulations 
and used to establish the low energy limits
of the binary collision approximation.

An extremely important issue in deploying molecular dynamics
to model collision processes in covalent and ionic
solids is how to incorporate
energy transfer mechanisms between electrons and ions \cite{Stoneham}.
A good description of dynamical processes 
in energetic collisions, such as
initial displacement damage, 
relaxation processes, and the cooling phase as the energy
dissipates into the ambient medium,
requires a theoretical framework that encompasses all interactions
between ion-ion, ion-electron, and their interaction with
the thermal surroundings. Especially, it
should capture the  nonequilibrium
thermodynamic nature of these physical processes
involving a wide range of energy scales, from 
a low energy electron-phonon interaction regime
to a high energy
radiation damage regime.
\cite{Allen,Flynn,Finnis,Johnson,Caro}. 
Traditional MD simulations can capture the thermal behavior
of an insulator. Since they 
do not take into account coupling
between the phonons and the conduction
electron system, obviously, these simulations
underestimate the heat-transfer rate for noninsulating materials.
In addition to  lattice thermal conductivity, 
the issue of the conductivity due to electrons must be addressed.
Furthermore, a correct description of the electronic stopping power
should be incorporated into MD simulations for high energy implantation.
For example,
in sputtering processes by particle bombardment,
examination  of
 MD simulations  with and
without inelastic electronic energy loss has
established
that, independent of the ion's mass or energy,
the inelastic electronic energy losses by target
atoms within the collision cascade  have greater influence on  the ejected
atom yield than the ion's electronic losses \cite{Jakas}. This is  in contrast
to the belief that the electronic loss mechanism is important only for
cascades initiated by light ions or by heavy ions at high bombardment energies
\cite{Jakas}. Although a convincing experimental verification of the
electronic effects in sputtering is still lacking, the effects should
be relevant to defect production rates, defect mobility and annealing, etc.\
\cite{Robinson-KDan,Nieminen-KDan}.
Also as shown in Ref.\ \onlinecite{Jensen-cover},
traditional MD simulations produce extremely long channeling
tails due to the absence of electronic stopping.
In order to incorporate the ion-electron interaction into
molecular dynamics simulations, a simple scheme was proposed by
adding a  phenomenological term,
which describes the inelastic electronic stopping in the high energy
radiation damage regime, while also capturing the thermal conductivity
by coupling low energy ions to a thermal reservoir \cite{Caro-Victoria}.
The empirical expression used in Ref. \onlinecite{Caro-Victoria} for 
the strength of the ion-electron coupling is a function of the local
electronic density.
At the low charge density limit,  a density
functional result was reproduced \cite{Echenique},
 and at the high charge density limit,
the linear response results were captured. 
In the same spirit,
we develop a stochastic MD model incorporating the electronic 
stopping power as a damping mechanism. Our model
is based on an effective charge theory \cite{BK} with the electronic
stopping power  factorized into two parts. One  is the effective charge
of the incident ion, which is a globally averaged quantity  determined
by the average unbound electron density in the medium.
The other factor is the electronic
stopping power for a proton, for which the  same local
density functional results are used.
Naturally, our damping mechanism incorporates both regimes, i.e., the electronic
stopping regime and the electron-phonon interaction regime, into
our molecular dynamics simulation, because the inelastic loss for
a proton  exhibits a similar density dependence as prescribed
in Ref. \onlinecite{Caro-Victoria}, with additional modifications due to
the velocity dependence of the effective charge. 
In the present work, however, we emphasize
mainly the electron stopping power in  the high energy regime ($\sim$ keV to 
$\sim$100keV), i.e., the electrons behave as an
 energy sink. The validity of the
model for
the electronic heat conduction regime will be discussed elsewhere.
In the following, for boron and arsenic implants into
single-crystal silicon in both the channeling  and off-axis
directions,
we will show that a classical MD with the
physically-based damping mechanism can generate
dopant profiles in excellent agreement with experimentally-measured
profiles obtained by secondary-ion mass spectrometry (SIMS).

As discussed above, the phenomenological model we have developed for electronic
stopping power is successfully implemented into both BCA Monte
Carlo programs and MD simulations. Wide applicability requires
that a model be valid for different implant species over a wide
range of energies. We emphasize that this electronic stopping model
is accurate both for boron and arsenic implants, thus providing
a crucial test of the generality and validity of the model in capturing
the correct physics of electronic stopping.

The paper is organized as follows. In Sec.\ II, we present the
phenomenological model for electronic stopping power in detail.
Atomic units $e= \hbar = m_{e} =1$ are used throughout the paper
unless  otherwise specified.
In Sec.\ III, we briefly discuss  different electronic stopping
models implemented on the versatile
BCA Monte Carlo simulation platform, MARLOWE \cite{RT,Eckstein}.
Then the results of the BCA Monte Carlo simulations on a rare-event
algorithm enhanced UT-MARLOWE
platform \cite{Yang-3.0} with our electronic stopping model are
summarized. In Sec.\ IV, the results of the MD with the inelastic
electronic energy loss are presented. In  Sec.\ V, we make closing
remarks and point out directions for future studies.

\section{The Model}
%In the local density approximation, i.e., each volume element $d\Omega$
%of the solid is assumed to an independent plasma,
%the final
%stopping power is evaluated by averaging over the whole space
%weighted by the local charge density distribution $\rho({\bf x})$:
%\begin{equation}
%S_{e} = \int S_{p}(v, r_{s}) [Z_{1}^{*}(v,r_{s})]^{2} \rho({\bf x})d\Omega
%\label{local_approximation}
%\end{equation}
%with $r_{s} = [3/(4\pi\rho({\bf x}))]^{1/3}$. Here $S_{p}(v, r_{s})$is
%the interaction of a particle of unit charge and velocity $v$ with a plasma
%of density $\rho({\bf x})$. $Z_{1}^{*}$ is effective charge of the 
%projectile.

According to the Brandt-Kitagawa (BK) theory \cite{BK}, the electronic stopping
power of an ion can be factorized into two components based on
an effective charge
scaling argument. One is the effective charge of the ion
(if not fully ionized), $Z_{1}^{*}$,
which is in general a function of ion velocity $v$ and the 
charge density of the target $\rho$, or equivalently, the one electron
radius $r_{s} = [3/(4\pi\rho({\bf x}))]^{1/3}$;
the other is the electronic stopping power for a proton, $S_{p}(v, r_{s})$.
In the local density approximation, therefore,
the total inelastic energy loss $\Delta E_{e}$ of an ion of 
constant velocity
$v$ 
is
\begin{equation}
\Delta E_{e} = \int [Z_{1}^{*}(v,r_{s})]^{2} S_{p}(v,r_{s}) dx,
\label{local_approximation}
\end{equation}
where the integral is along the ion path.
Since the effective charge is a continuous function of 
electronic density,
mathematically, it is always possible to find a mean
value, $r_{s}^{\circ}$ of $r_{s}$, such that Eq.\ (\ref{local_approximation})
can be rewritten as
\begin{equation}
\Delta E_{e} =[ Z_{1}^{*}(v,r_{s}^{\circ})]^{2} \int S_{p}(v,r_{s})
dx.
\label{mean}
\end{equation}
If the effective charge is a  slowly varying function
of  space, physically, 
this means that
$r_{s}^{\circ}$ describes an average number of unbound
electrons in the sea and thus 
can be assumed to determine 
the Fermi surface. Therefore, we have the relation between
the Fermi velocity and $r_{s}^{\circ}$:
\begin{equation}
v_{F} = \frac{1}{\alpha r_{s}^{\circ}},
\label{Fermi_velocity}
\end{equation}
where $\alpha =[4/(9\pi)]^{1/3}$.
We note that this $r_{s}^{\circ}$ will be the only tunable parameter in
our electronic stopping power model.

Next we turn to a simple statistical model for this partially
ionized, moving projectile.
For an ion with $N= Z_{1} - Q$ bound electrons, where
$Q$ is the charge number of the ion of atomic number $Z_{1}$, a radially
symmetric charge density
\begin{equation}
\rho_{e} = \frac{N }{4\pi \Lambda^2 r} \exp\left(-\frac{r}{\Lambda}\right)
\end{equation}
is used in the BK theory.
Here $\Lambda$ is the ion size parameter,  a function of
the fractional ionization,
$q= (Z_{1}- N)/Z_{1}$. The total energy of the electrons comes
from the sum of
the kinetic energy estimated by the local density approximation,
the electron-electron interaction in the Hartree approximation
weighted by a variational parameter $\lambda$ to account for correlation,
and the Coulomb energy of the electrons in the electric field of the nucleus.
A variational approach minimizing the total energy leads to the
following dependence of the ion size on the ionization fraction $q$:
\begin{equation}
\Lambda = \frac{2 a_{0} (1-q)^{2/3}}{Z_{1}^{1/3}\left[1-(1-q)/7\right]},
\end{equation}
where $a_{0} = 0.24005$.
In the BK theory, 
the generalized Lindhard theory of the electronic stopping
in a homogeneous electron gas with an electron density
$n = 3/(4\pi \bar{r}_{s}^{3})$ is used.  The total electronic stopping
is
estimated from the sum of the energy loss in soft, distant collisions, 
i.e., small momentum transfers with target electrons seeing a charge
$qZ_{1}$, and the energy loss to the target electrons 
 experiencing increased nuclear interaction
 in hard, close
collisions corresponding to large momentum transfers.
As extensively discussed in the literature  
(see, e.g., Ref.\ \onlinecite{BK,Kreussler,ZBL},
and references therein),
it is assumed that the charge state of a  proton in a solid
is unity.
Given an ionization fraction $q$ and
using the scaling argument for the ratio of ion stopping to the proton
stopping at the same velocity,
the BK theory produces a simple
expression for the fractional effective charge of an ion \cite{BK,ZBL}
\begin{equation}
\gamma(\bar{r}_{s})
 = q + C(\bar{r}_s)(1-q)
\ln\left[1 + \left(\frac{4\Lambda}{\bar{r}_{s}}\right)^{2}
\right],
\label{effective_charge}
\end{equation}
where $C(\bar{r}_{s})$ is weakly dependent on the target and has
a numerical
value of about $1/2$. 
We will set $C = 0.5$  below.
 Then, the  effective charge is
\begin{equation}
Z_{1}^{*} = Z_{1} \gamma(\bar{r}_s).
\end{equation}
For our model, using the procedure (\ref{mean}) outlined above, 
this dependence of $\bar{r}_{s}$ is identified with
the dependence of the mean value $r_{s}^{\circ}$. Therefore,
the effective charge $Z_{1}^{*}$ has a nonlocal, i.e.,
spatially  independent,  character and depends
on the Fermi surface.
In the above discussion, as can be seen,
$q$ is a parameter which is
not fixed by the BK theory.
For obtaining this ionization fraction, there
are velocity  and energy criteria originally proposed
by Bohr
\cite{Bohr} and Lamb \cite{Lamb}, respectively. 
Kitagawa also
used a statistical argument to justify scaling analyses
in terms of the scaling parameter $v_{1}/(v_{B}Z_{1}^{2/3})$ \cite{Kitagawa}.
Recently, the issue of which stripping criterion can give rise to
a better physical understanding has been raised \cite{Mathar}.
However, in light of the
 large amount of experimental data employed in Ref. \onlinecite{ZBL}
to extract an ionization scaling consistent with the Brandt-Kitagawa
theory, we will use this empirically verified scaling in our
model. As  summarized in Ref. \onlinecite{ZBL},
a new criterion  in the BK approach is proposed
\cite{BK,Kreussler}, i.e.,
a relative velocity criterion, which assumes that the electrons of the
ion which
have an orbital velocity lower than the relative velocity between the ion
and the electrons in the medium are stripped off. 
The relative velocity $v_{r}$
is obtained by averaging over the difference between the ion velocity
${\bf v}_{1}$ and the electron velocity ${\bf v}_{e}$ under the assumption
that the conduction electrons are a free electron gas in the ground
state, therefore,
 whose velocity distribution is isotropic.
Performing a further averaging of ${\bf v}_{e}$ over the Fermi sphere
leads to \cite{Kreussler}
\begin{eqnarray}
v_{r} & = & v_{1}\left(1 + \frac{v_{F}^2}{5 v_{1}^2}\right) 
\mbox{\hspace{22mm}for $v_{1} \geq v_{F}$,} \\
v_{r} & = & \frac{3 v_{F}}{4} \left(1 + \frac{2 v_{1}^{2}}{3v_{F}^2}
-\frac{v_{1}^4}{15 v_{F}^4}\right)
\mbox{\hspace{3mm}for $v_{1} < v_{F}$. }
\end{eqnarray}
For the ionization scaling, a form of the Northcliffe 
type\cite{Northcliffe} is then
assumed  for the scaling variable, i.e., the reduced relative velocity:
\begin{equation}
y_{r} = \frac{v_{r}}{v_{B}Z_{1}^{2/3}},
\end{equation}
where $v_{B}$ is the Bohr velocity and $v_{B}=1$ in our units.
The extensive experimental data for ions $3 \leq Z_{1} \leq 92$
are used in Ref. \onlinecite{ZBL} to determine
\begin{equation}
q = 1 - \exp [ -0.95 (y_{r} -0.07)].
\label{ionization}
\end{equation}
In Ref. \onlinecite{ZBL}, an ionization scaling fit
with
even tighter bunching of the experimental data along the fit is presented.
However, this approach entails a much more involved computational procedure
\cite{ZBL}. The accuracy level of Eq.\ (\ref{ionization})
is adequate for our present purposes.

In our model, the electronic stopping power for a proton is derived
from a nonlinear density-functional formalism \cite{Echenique}.
In the linear response theory, 
the energy loss per unit path length of a proton moving at
velocity $v$ in the electron gas is obtained by Ritchie \cite{Ritchie}
\begin{equation}
\left(\frac{dE}{dx}\right)_{\rm R} = \frac{2 v}{3 \pi}\left[\ln\left(1+\frac{\pi}
{\alpha r_{s}}\right) - \frac{1}{1+\alpha r_{s}/\pi}\right],
\label{p-stopping}
\end{equation}
using an approximation to the full random-phase approximation
dielectric function, which amounts to the exponential screening
potential around the ion induced
by density fluctuations of the electrons. The nonlinear,
density-functional calculation based on the formalism of
Hohenberg and Kohn, and Kohn and Sham \cite{Hohenberg,Kohn} has
been performed \cite{Echenique,ENAR,Echenique-Arnau} to obtain
the charge density and scattering
phase shifts 
for the conduction band as a function of energy
self-consistently. The final
stopping power for a proton is obtained via
\begin{equation}
\frac{dE}{dx} = \frac{3 v}{ k_{F} r_{s}^3}
\sum_{l=0}^{\infty}(l+1)\sin^{2}\left[\delta_{l}(E_{F})
-\delta_{l+1}(E_{F})\right],
\end{equation}
where $\delta_{l}(E_{F})$ is  the phase shift at the
Fermi energy for scattering of an electron of angular momentum
$l$  and $k_{F}$ is the Fermi momentum \cite{Ferrell}.
As shown in Ref. \onlinecite{ZBL,Mann,Brandt82},
comparison with expermental 
data demonstrates that the density functional treatment
provides an improvement over the linear response (dielectric)
 result \cite{LW}, which underestimates
the stopping powers. In our implementation, the result
of the nonlinear,
density functional formalism for the electronic stopping power
for a proton is used, which can be expressed as
\begin{equation}
S_{p}(v,r_{s}) = - \left(\frac{dE}{dx}\right)_{\rm R} G(r_{s}),
\end{equation}
where, for computational convenience, the correction factor $G(r_{s})$ takes
the form
\begin{equation}
G(r_{s})\! = \!1.00 + 0.717 r_{s} \!- 0.125 r_{s}^2\! -0.0124 r_{s}^{3}
\!+ 0.00212 r_{s}^{4}
\label{G}
\end{equation}
for $r_{s} < 6$. We note that a different correction factor was
used in Refs.\ \onlinecite{Azziz,Klein},  which does not have 
the following desired
behavior for $r_{s} \ll 1$. Since the density functional result
converges to the Ritchie formula as $r_{s}$ decreases towards
values sufficiently small compared to unity \cite{Echenique}, 
this requires that
the correction factor smoothly tend to unity as $r_{s} \rightarrow 0$.
Obviously, the above $G(r_{s})$ possesses the correct convergence
property.

The last ingredient needed for our model is the charge distribution
$\rho({\bf x})$
for silicon atoms in the crystal. We use the solid-state 
Hartree-Fock atomic charge
distribution \cite{ZBL}, which is spherically symmetric due to
the muffin-tin construction. In this approximation,
there is about one electron charge unit (0.798 electrons for Si)
left outside the muffin-tin. This small amount of charge can be
either distributed in the volume between the spherical atoms, resulting in
an interstitial background charge density $0.119 e/{\rm \AA}^{3}$, or
distributed between the maximal collision distance used in Monte
Carlo simulations and the muffin-tin radius (see details below).

\section{BCA Monte Carlo simulation results}
First, in 
comparison with other electronic stopping models used in Monte
Carlo simulations based on the MARLOWE platform, 
we stress that in our model the effective charge is a nonlocal
quantity, neither explicitly dependent on the impact parameter
nor on the charge distribution,
and the stopping power for a proton depends on
the local charge density of the solid. 
A purely nonlocal version of the BK
theory was implemented into MARLOWE \cite{Azziz},
in which both the effective charge and the stopping power
for a proton depend on a single nonlocal parameter, namely, the
averaged one electron radius.
Its results 
demonstrated that energy loss for well-channeled ions in the keV
region has high sensitivity to the one-electron radius in the channel.
It was pointed out that a correct density distribution is needed to
account for
the electronic stopping in the channel
\cite{Murthy,Azziz}. Later, a purely local version of
the BK theory was developed to take into
account the charge distribution of the electrons \cite{Klein,Klein_APL}.
Comparison with other electronic stopping models, such as Lindhard
and Scharff \cite{LS}, Firsov \cite{Firsov}, and the above nonlocal
implementation \cite{Azziz}, showed a marked improvement in modeling
electronic stopping in the channel \cite{Yang,Klein,Klein_APL}.
 Good agreement between
simulated dopant profiles and the SIMS profiles for boron implants
into $\langle 100 \rangle$ single crystal silicon was obtained. However, 
this purely local implementation of the BK theory
did not successfully model the electronic stopping 
 for the boron implants
into the $\langle 110 \rangle$ axial channel and arsenic implants 
as noted in Ref.\ \onlinecite{Yang}. 

In the present work,
UT-MARLOWE \cite{Yang-3.0} was selected as the platform for our
electronic stopping model implementation. UT-MARLOWE is an extension
of the MARLOWE code for simulating the behavior of energetic ions in
crystalline materials \cite{RT,Eckstein}. It has been enhanced with:
(i) atomic pair-specific interatomic potentials for $B$-$Si$, $B$-$O$,
$As$-$Si$,
$As$-$O$ 
for nuclear stoppings \cite{ZBL},
(ii) variance reduction algorithm implemented for rare events,
(iii) important implant parameters accounted for, e.g., tilt and rotation angles,
the thickness of the native
oxide layers, beam divergence, and wafer temperature, etc.
In our simulations, we have turned off certain options, such as
the cumulative damage model in the UT-MARLOWE code, which is
a phenomenological model to estimate defect production and recombination
rates. Individual ion 
trajectories were simulated under the BCA and the overlapping of
the damage caused by different individual cascades was
 neglected. In order to test 
the electronic stopping model we also used  low dose ($10^{13}/{\rm cm}^{2}$)
implants 
so that cummulative damage effects do not
significantly complicate dopant profiles
\cite{Yang}. Also, for the simulation
results we report below, 16${\rm \AA}$ native oxide surface layer,
300K wafer temperature
were used. The maximum distance for searching a collision
partner is 0.35 lattice constant, the default value in the UT-MARLOWE 
\cite{Yang-thesis}. The excess  charge outside the muffin-tins
is distributed in the space between this maximum collision distance
and the muffin-tin radius. In the simulation,
the electronic stopping power is evaluated continuously
along the path the ion traverses through regions of varying
charge density, i.e., the energy loss is given by
\begin{equation}
\Delta E_{e} =  \int_{{\rm ion\,\,\, path}} [Z_{1} \gamma(v_{1},r_{s}^{\circ})]^2 
 S_{p}(v_{1}, r_{s}({\bf x}))dx.
\label{e-stop}
\end{equation}

In the simulations, 
the free parameter $r_{s}^{\circ}$ was adjusted to yield
 the best results in overall
comparison with the 
experimental data. The value $r_{s}^{\circ} =1.109{\rm \AA}$
was used for both
boron and arsenic ions for all energies and incident directions.
This value is physically reasonable for silicon.
Note that  the unbound electronic
density in silicon with only valence electrons
taken into account will give rise to a value of $1.061 {\rm \AA}$ for
$r_{s}$. The fact that our $r_{s}^{\circ}$ value is
greater than $1.061 {\rm \AA}$ indicates that
not all valence electrons participate in stopping
the ion as unbound electrons.

We display the Monte Carlo dopant profile simulation results as follows.
We note in passing that the lower and upper limits
of energy used in our simulations are determined by the energy range
of the SIMS data available to us.

\begin{figure}
\vbox{\vspace{-0.80 in} \hspace{1.5 in}
\includegraphics{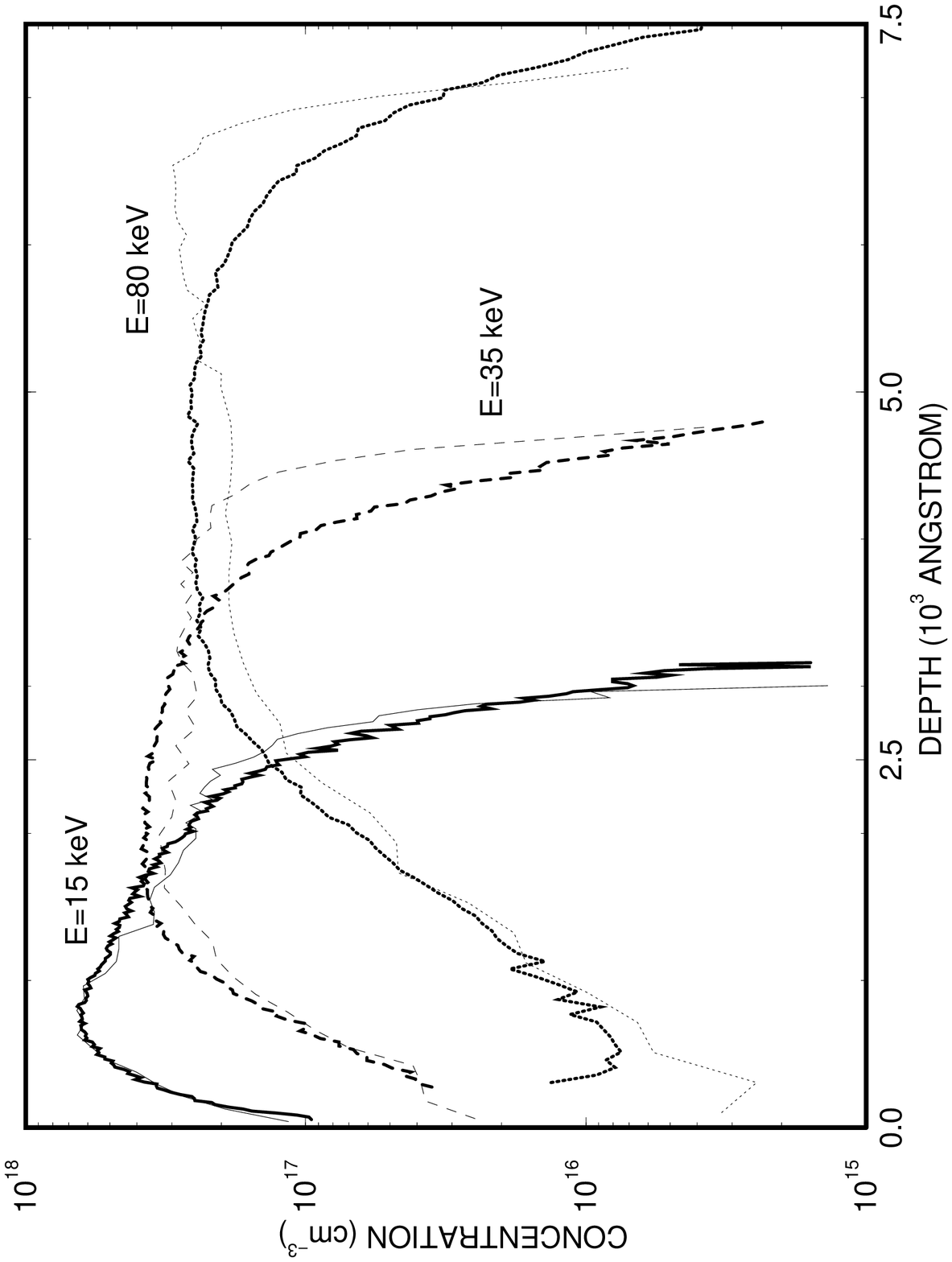}
\vspace{2.80 in}}

\caption[]{Boron implant profiles for the $\langle 100 \rangle$
direction with energies ranging from 15keV to 80keV. Zero tilt and
rotation angles. The thick lines are SIMS data.
}
\label{fig-b100}
\end{figure}

\begin{figure}
\vbox{\vspace{-0.80 in} \hspace{1.5 in}
\includegraphics{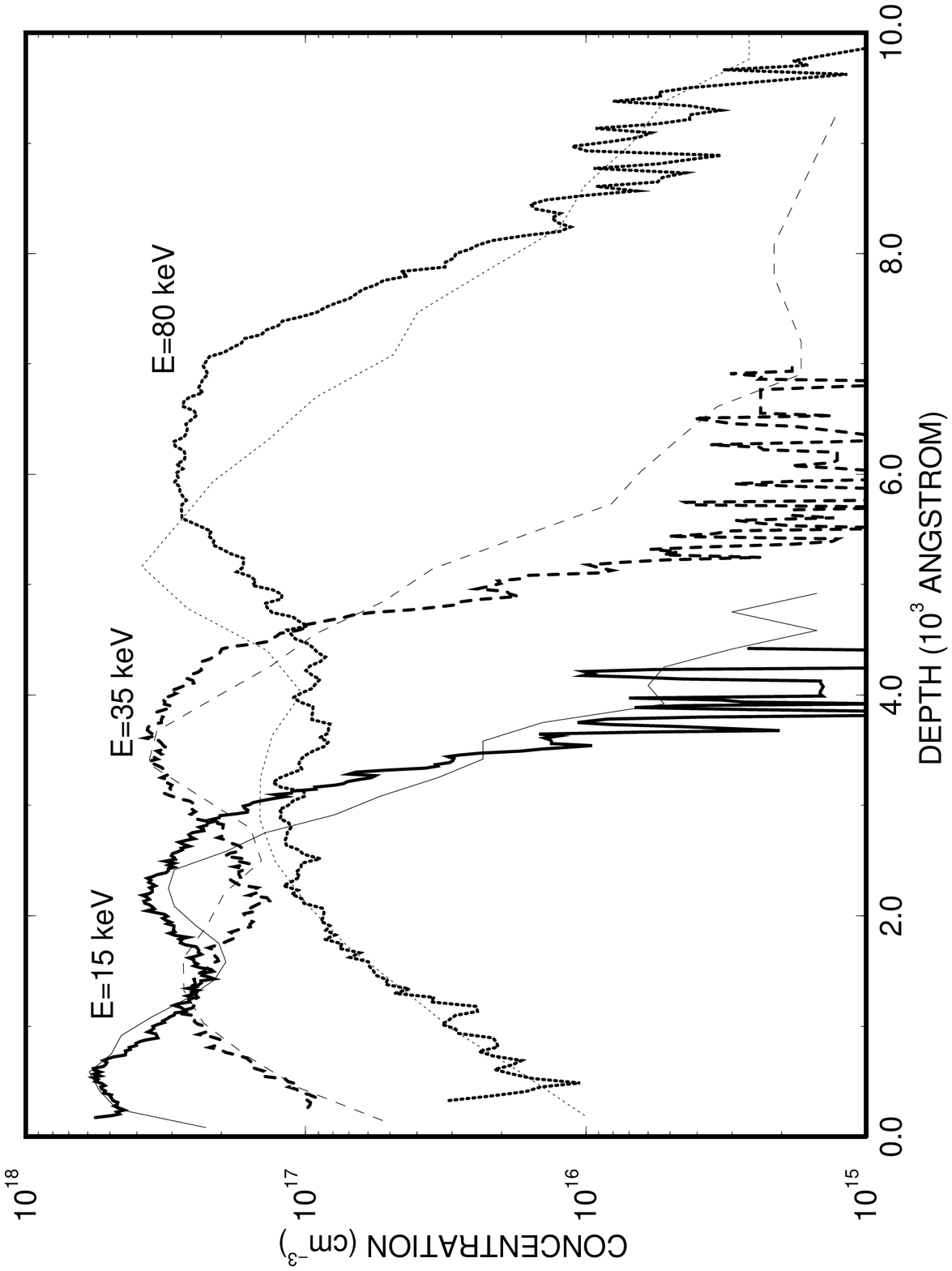}
\vspace{2.80 in}}

\caption[]{Boron implant profiles for the $\langle 110 \rangle $
direction with energies ranging from 15keV to 80keV. Zero tilt and
rotation angles. The thick lines are SIMS data.
}
\label{fig-b110}
\end{figure}

\begin{figure}
\vbox{\vspace{-0.80 in} \hspace{1.5 in}
\includegraphics{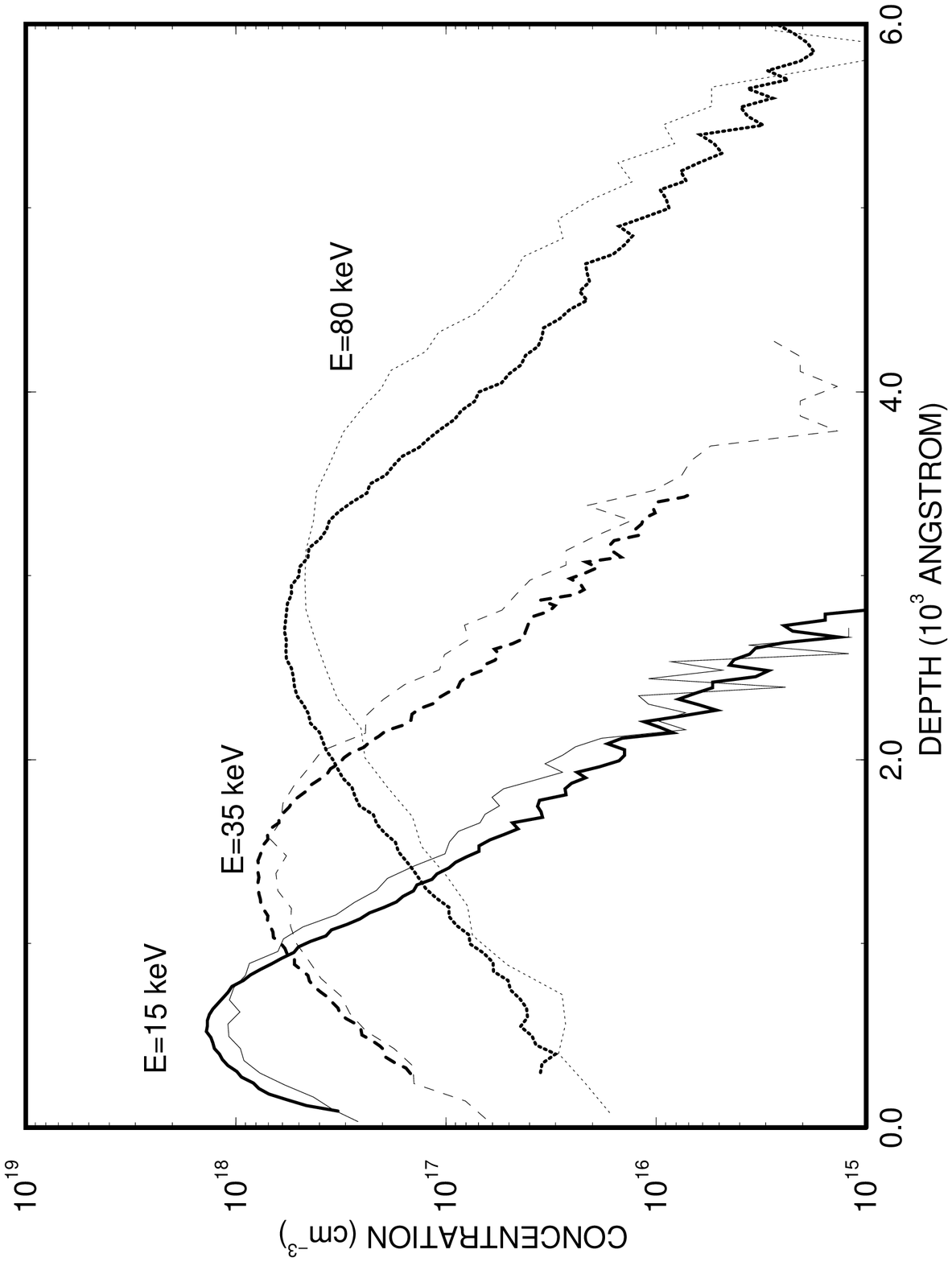}
\vspace{2.80 in}}

\caption[]{Boron implant profiles for the $\langle 100 \rangle$ direction
with the tilt $=7^{\circ}$ and rotation $=30^{\circ}$. The thick lines
are SIMS data.
}
\label{fig-b730}
\end{figure}

In Figs.\ \ref{fig-b100}, \ref{fig-b110} and \ref{fig-b730}, we show
boron dopant profiles for the
energies 15keV, 35keV, and 80 keV along $\langle 100 \rangle$,
$\langle 110 \rangle$, and the off-axis direction with
tilt $=7^{\circ}$ and rotation $=30^{\circ}$, respectively.
It can be seen that the overall agreement with the SIMS data
is excellent.
In Fig.\ \ref{fig-b100}, 
the
simulations show a good fit for the cutoff range. In the high energy
regime, the simulated distribution shows a slightly peaked structure.
This can be attributed to a strong channeling due to insufficient
scatterings in the implants. We have noticed that by increasing, e.g.,
the native oxide layer thickness, the peak can be reduced. 
For the $\langle 110 \rangle$ channeling case,
the distribution indicates a possibility that
the total electronic stopping power
along the channel is a little too strong at the high energy end.
However, it should be kept in mind that for this
channel, the UT-MARLOWE 
model becomes sensitive to the 
multiple collision parameter which is employed
as an approximate numerical correction for the effect
of multiple overlapping nuclear encounters.
It is not clear how to
separate the contributions from  these two different sources.

\begin{figure}
\vbox{\vspace{-0.80 in} \hspace{1.5 in}
\includegraphics{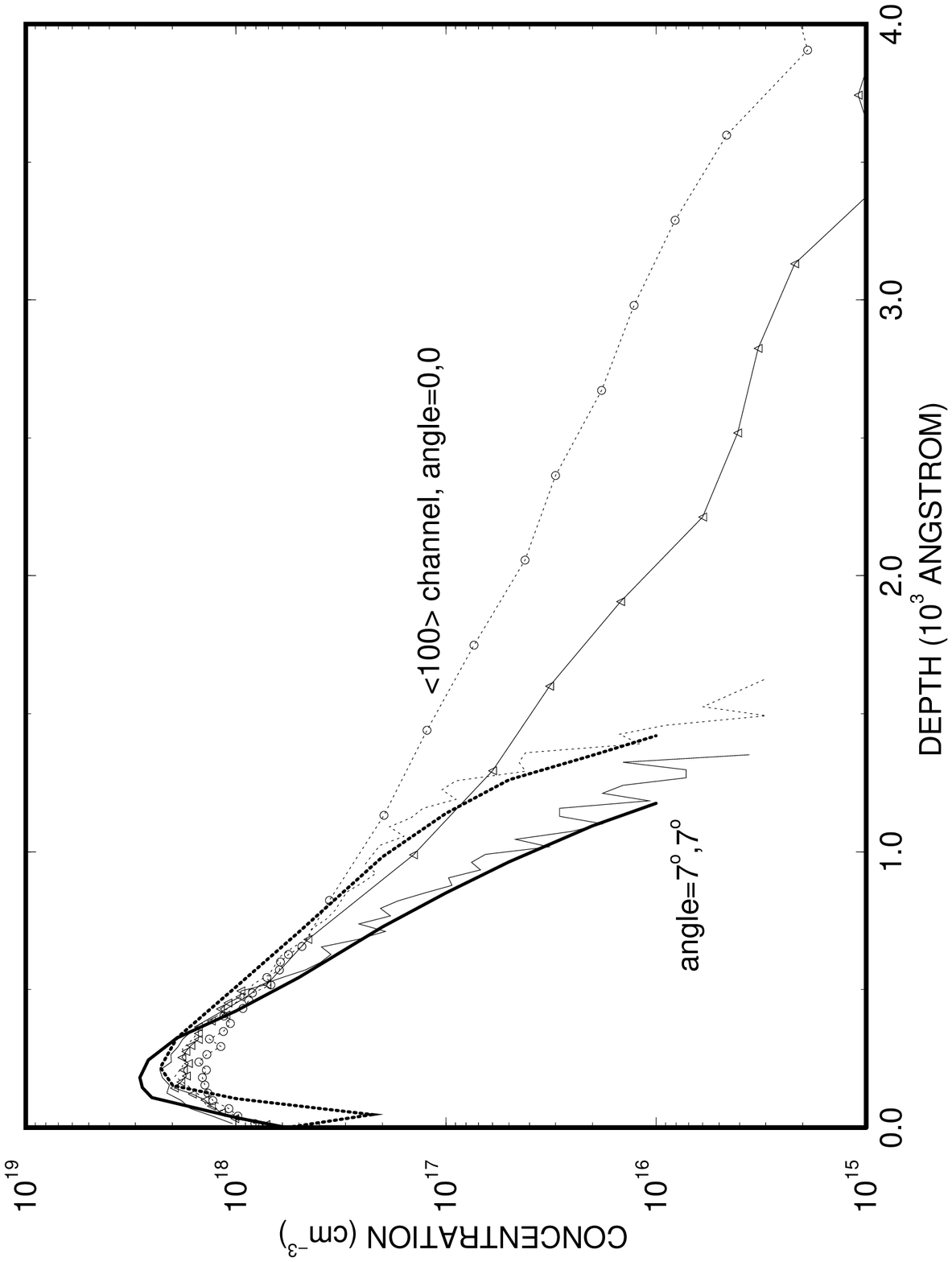}
\vspace{2.80 in}}

\caption[]{Boron implant profiles for the $\langle 100 \rangle$ channeling
and for the off-axis direction
with the tilt $=7^{\circ}$ and rotation $=7^{\circ}$. The implant energy is 5keV.
Thick lines: SIMS data; Thin lines without symbols: BCA Monte Carlo simulation
with full electronic stopping power;
Circles: BCA Monte Carlo simulation with the artificially reduced electronic
stopping power for the channeling case; Triangles: the corresponding
case for the off-axis direction (see text).
}
\label{fig-b77}
\end{figure}

For comparison, in Fig.\ \ref{fig-b77}, we also display 
 a low energy (5keV) implant
case \cite{xxx}. Again the agreement for both the channeling and off-axis
directions is striking (thin lines without symbols).
 In order to illustrate the
importance of electronic stopping power at this low energy for boron
implants, we have used an artificially reduced electronic
stopping power, i.e, multiplying $\Delta E_{e}$ in Eq.\ (\ref{e-stop})
by a factor of $1/10$ in the simulation, to genearte 
dopant profiles in the channeling and off-axis directions.
From Fig.\ \ref{fig-b77}, evidently, it can be concluded that,
for boron implants even in this low energy regime, electronic
stopping power has  significant influence on the channeled tail of the
dopant
distribution and on the cutoff range
for both channeling and off-axis directions.

\begin{figure}
\vbox{\vspace{-0.80 in} \hspace{1.5 in}
\includegraphics{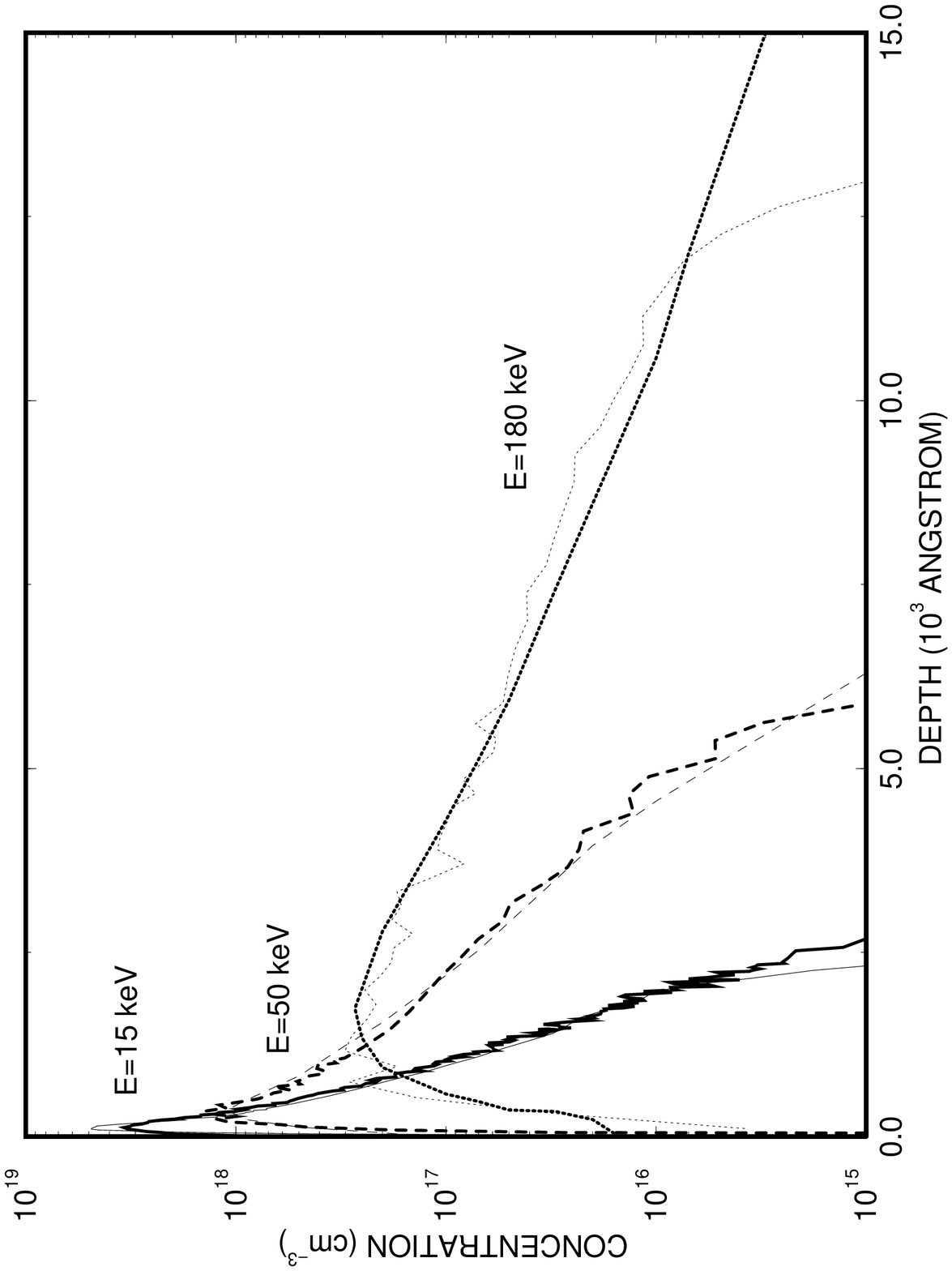}
\vspace{2.80 in}}

\caption[]{Arsenic implant profiles for the $\langle 100 \rangle$
direction with energies ranging from 15keV to 180keV. Zero tilt and
rotation angles. The thick lines are SIMS data.
}
\label{fig-a100}
\end{figure}

\begin{figure}
\vbox{\vspace{-0.80 in} \hspace{1.5 in}
\includegraphics{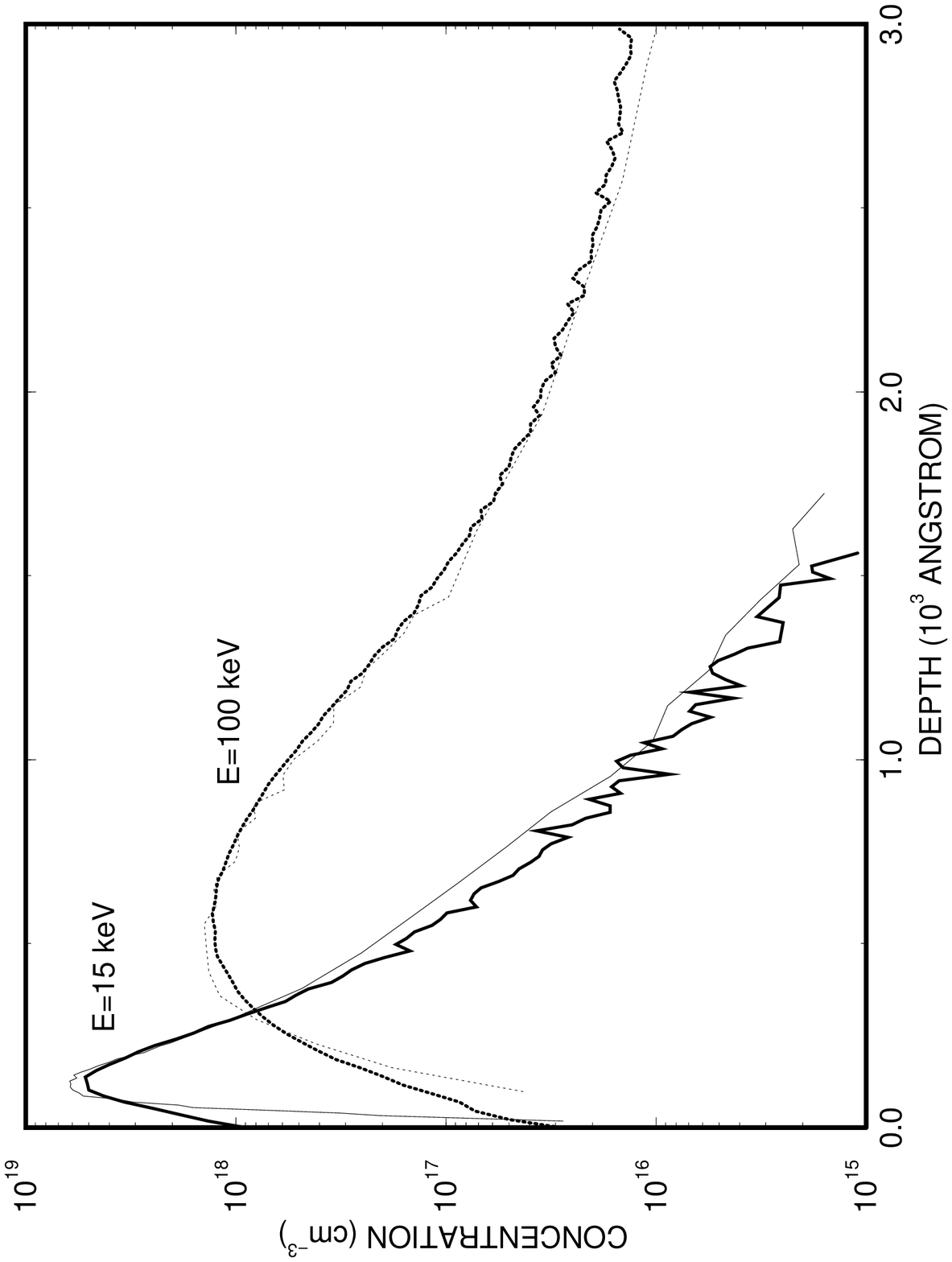}
\vspace{2.80 in}}

\caption[]{Arsenic implant profiles for the $\langle 100 \rangle$ direction
with the tilt $=8^{\circ}$ and rotation $=30^{\circ}$. The thick lines
are SIMS data.
}
\label{fig-a830}
\end{figure}

\begin{figure}
\vbox{\vspace{-0.80 in} \hspace{1.5 in}
\includegraphics{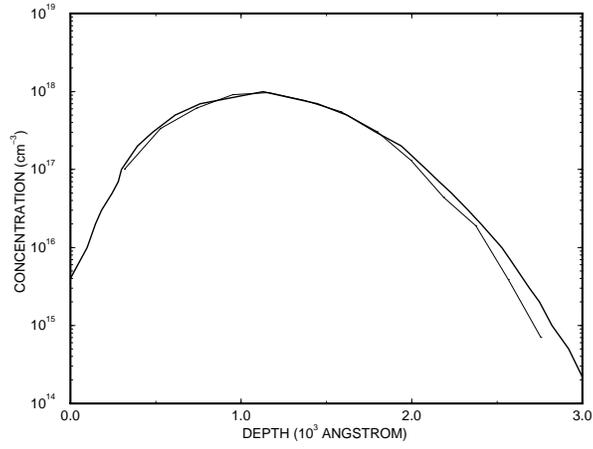}
\vspace{2.80 in}}

\caption[]{Arsenic implant profile into amorphous silicon with the 
implant energy being 180keV.
The thick line
is SIMS data.
}
\label{fig-amo}
\end{figure}

Figs.\ \ref{fig-a100} and \ref{fig-a830} show
arsenic dopant profiles for 
energies ranging from 15keV to 180 keV along $\langle 100 \rangle$,
 and the off-axis direction with
tilt $=8^{\circ}$ and rotation $=30^{\circ}$, respectively. It can readily
be concluded that our electronic model works successfully with
arsenic as well as boron implants into crystalline silicon. 
For comparison,
a case of arsenic implant into amorphous silicon is also shown in
Fig.\ \ref{fig-amo}. 
The implant energy is $180$keV.
The effect
of electronic stopping, which shows clearly in the long sloping
channeling
tail in the crystalline counterpart (see Fig.\ \ref{fig-a100}), 
is less prominent
for the amorphous case (Fig.\ \ref{fig-amo}).

\begin{figure}
\vbox{\vspace{-0.80 in} \hspace{1.5 in}
\includegraphics{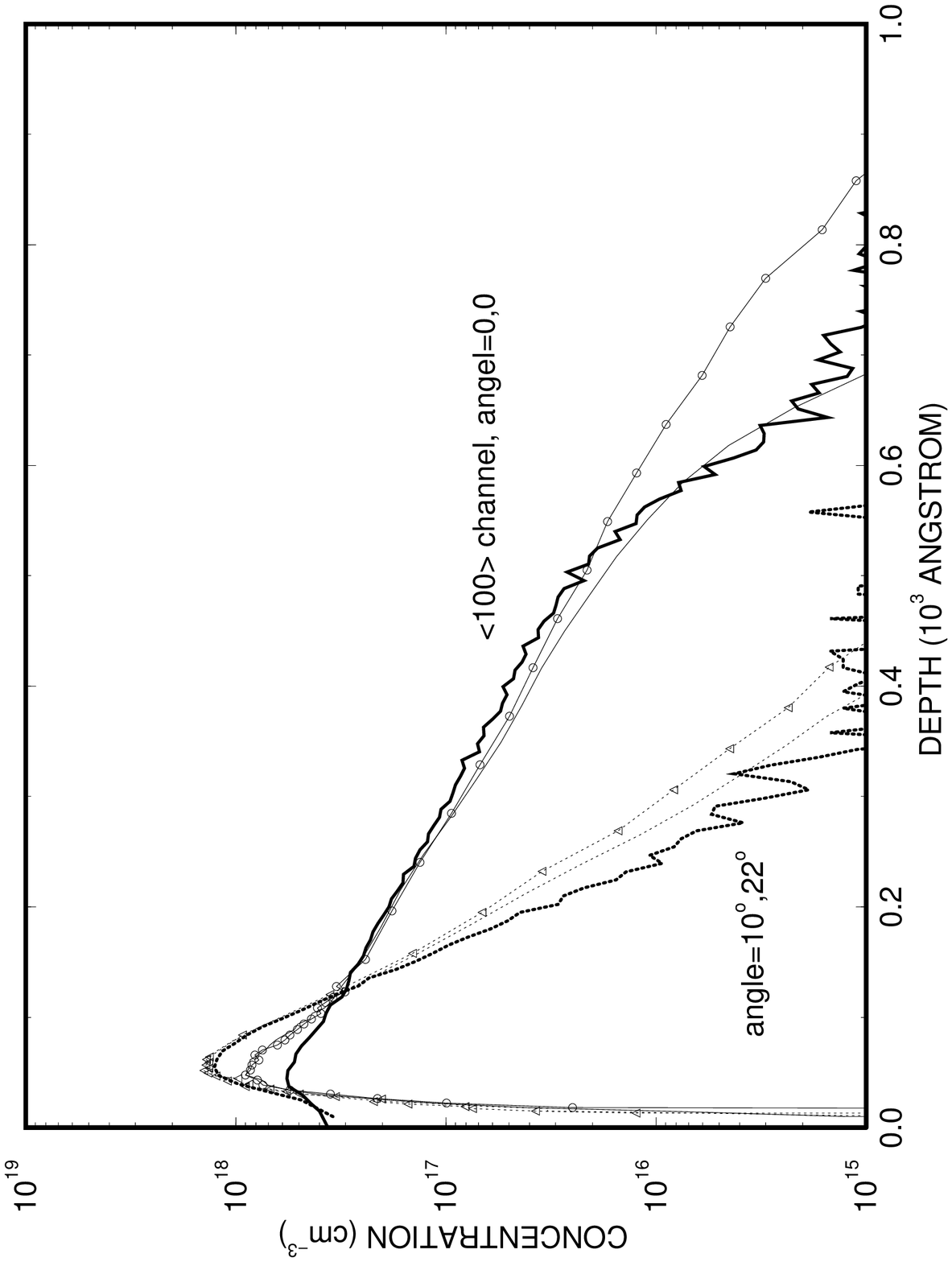}
\vspace{2.80 in}}

\caption[]{Arsenic implant profiles for the $\langle 100 \rangle$ channeling
and for the off-axis direction
with the tilt $=10^{\circ}$ and rotation $=22^{\circ}$. 
The implant energy is 5keV.
Thick lines: SIMS data; Thin lines without symbols: BCA Monte Carlo simulation
with full electronic stopping power;
Circles: BCA Monte Carlo simulation with the artificially reduced electronic
stopping power for the channeling case; Triangles: the corresponding
case for the off-axis direction (see text).
}
\label{fig-a5}
\end{figure}

To examine the role that electronic
stopping power has on arsenic implants in the 
low energy regime ($\sim 5$keV), 
we again simulated arsenic dopant profiles with the artificially
reduced
electronic stopping power. For these low energy implants,
the oxide layer thickness $3{\rm \AA}$ were used in the
BCA simulations on account of the fact that the wafers used
for these implants were treated by dilute HF etch for 30 seconds, then
implanted within 2 hours to prevent native oxide regrowth \cite{Tasch}.
In Fig.\ \ref{fig-a5},
we show that our electronic stopping power model
is successful in
both the $\langle 100 \rangle$ channeling and off-axis
directions (thin lines without symbols). 
Clearly, the artificial reduction of electronic
stopping power leads to incorrect dopant distributions
and cutoff ranges for both the channeling and off-axis
directions, although the deviations indicate
a less
significant contribution from electronic stopping
 for arsenic implants than for boron implants
at the  energy $5$keV.
However, the deviation in the cutoff range
due to the 
electronic stopping power reduction
for the channeling
case is still 
significant. Obviously, this reinforces
the conclusion that, for channeling implants
even at  low energies, electronic stopping
is not negligible.

In summary, the above results 
 demonstrate clearly that 
our electronic stopping power indeed captures the correct physics
of the electronic stopping for ion implants into silicon
over a wide range of implant energies.

\section{ MD simulation results}
\begin{figure}
\vbox{\vspace{-0.80 in} \hspace{1.5 in}
\includegraphics{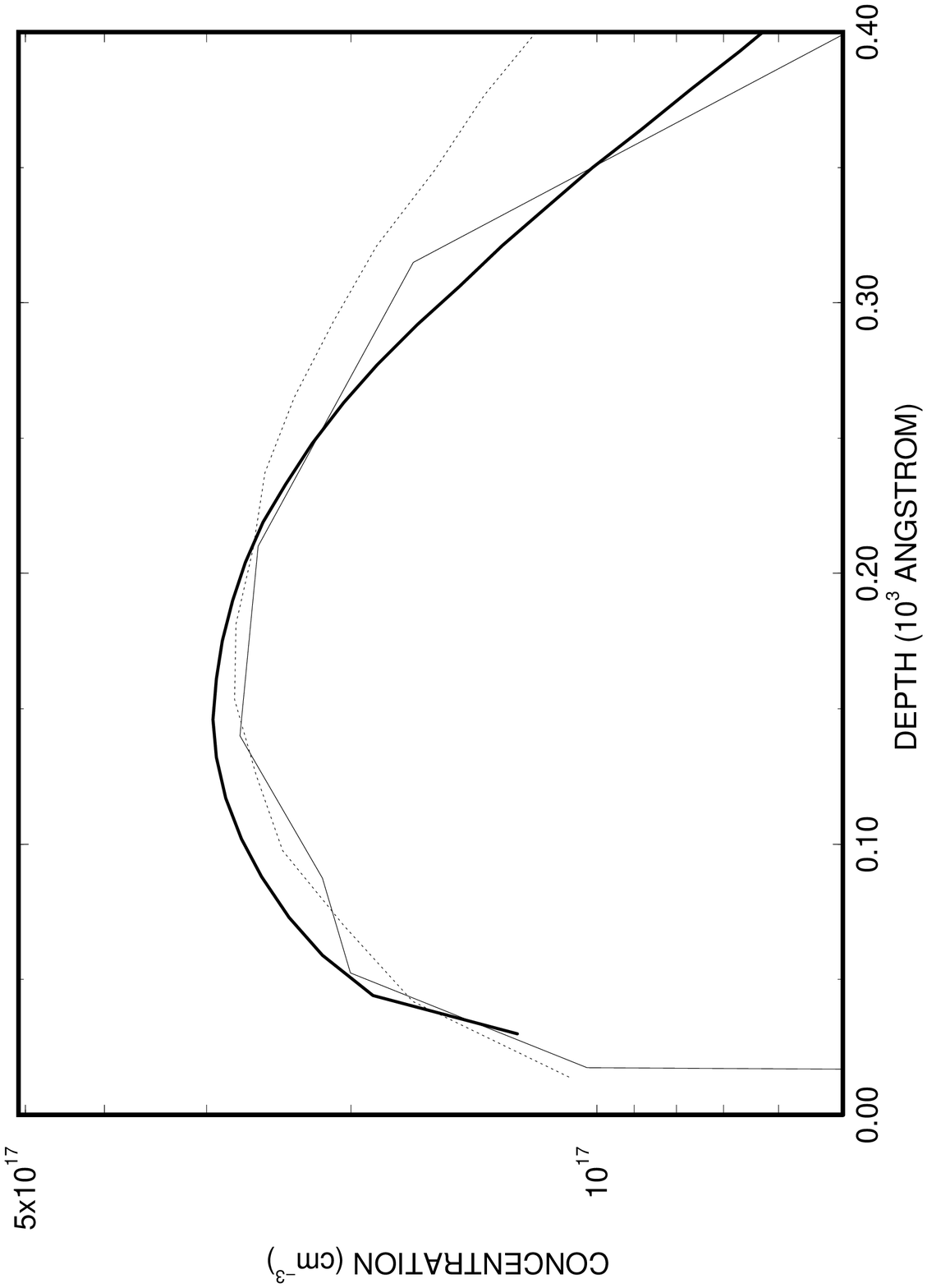}
\vspace{2.80 in}}

\caption[]{Molecular dynamics simulation: boron implant profiles
into the $\langle 100 \rangle$ channel with tilt $=10^{\circ}$,
rotation $=22^{\circ}$. Implant energy is 5keV.
Thick line: SIMS data. Thin line: MD simulation. Dotted line:
BCA Monte Carlo simulation.
}
\label{fig-mdb5}
\end{figure}

\begin{figure}
\vbox{\vspace{-0.80 in} \hspace{1.5 in}
\includegraphics{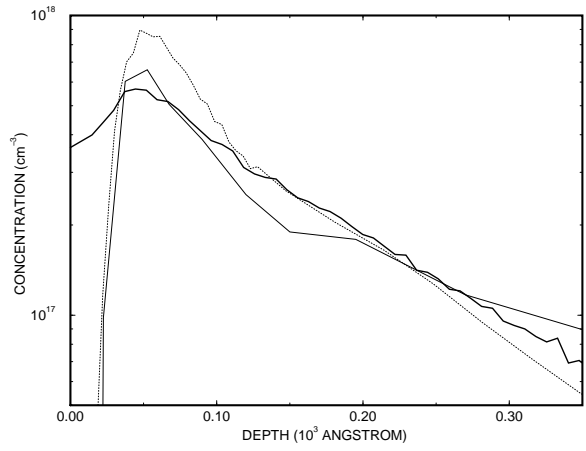}
\vspace{2.80 in}}

\caption[]{Molecular dynamics simulation: arsenic implant profiles
into the $\langle 100 \rangle$ channel with zero tilt,
rotation angles. Implant energy is 5keV.
Thick line: SIMS data. Thin line: MD simulation. Dotted line:
BCA Monte Carlo simulation.
}
\label{fig-mda500}
\end{figure}

\begin{figure}
\vbox{\vspace{-0.80 in} \hspace{1.5 in}
\includegraphics{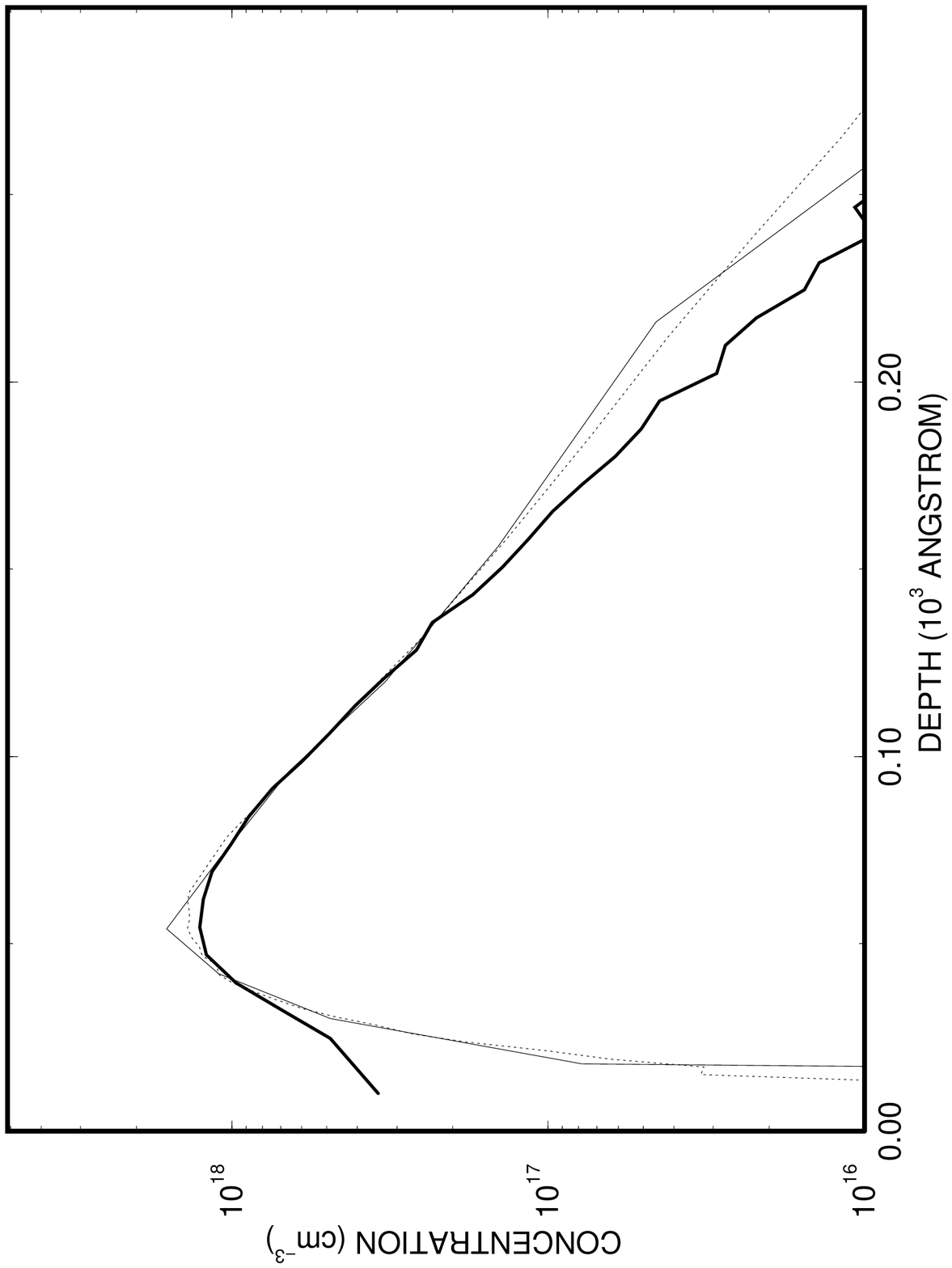}
\vspace{2.80 in}}

\caption[]{Molecular dynamics simulation: arsenic implant profiles
into the $\langle 100 \rangle$ channel with tilt $=10^{\circ}$,
rotation $=22^{\circ}$. Implant energy is 5keV.
Thick line: SIMS data. Thin line: MD simulation. Dotted line:
BCA Monte Carlo simulation.
}
\label{fig-mda51022}
\end{figure}

We have also used  classical  molecular dynamics simulation
to study the electronic stopping power as one of the damping mechanisms
in the high energy regime,
as discussed above. 
Here we demonstrate that experimental data, such as
SIMS, can be used to test the validity of
this physically-based damping model.
The interaction between silicon atoms are modeled
by 
Tersoff's empirical potential \cite{Tersoff}:
\begin{equation}
E= \frac{1}{2} \sum_{i\neq j} f(r_{ij})
\left[V_{R}(r_{ij})-b_{ij}V_{A}(r_{ij})\right],
\end{equation}
where $f(r_{ij})$ is a cutoff function that restricts interactions
to nearest neighbors, $V_{R}(r_{ij})$ and $V_{A}(r_{ij})$ are
pair terms, and $b_{ij}$ is a many-body function that can be regarded
as an effective Pauling bond order. 
We have  modified the  repulsive part of the Tersoff potential
by splinning to the
 ZBL universal potential at close-range.
\cite{ZBL}.
The ZBL universal
potential is also used to model the ion-silicon interactions.
In our full MD simulations for the low dose implantation, 
the lattice temperature was initialized to 300K 
and
the above electronic stopping model was applied to all the atoms.
The only modification required for implementation in MD is to take
into account the contributions from multiple silicon atoms to
the local electron density, while ensuring that the background electron
density is only counted once.
For each individual cascade, all recoils and the accumulation of damage
in  the ion path are  taken into account.
 Using the parameter value $1.109$
for $r_{s}^{\circ}$ from the comparison of BCA Monte Carlo simulation
results with the SIMS data, we
have simulated the implantation of low energy boron and arsenic
ions into the $Si$ $\{100\}(2\times1)$ surface at energies between 0.5keV
and 5keV, with both channeling and off-axis directions of
incidence. 
We mention
here that, for the $\langle 100 \rangle $ channeling case up to
0.16 keV ($32 \%$) of 0.5 keV boron implant energy  and
0.64 keV ($13 \%$) of 5keV arsenic implant energy
are lost via 
electronic stopping in our simulations. Simulations
were terminated when the total energy of the ion became less
than 5eV, giving typical simulation times of around 0.2ps.
Figs.\ \ref{fig-mdb5}, \ref{fig-mda500} and
\ref{fig-mda51022} show the calculated dopant concentration
profile for various energies and directions.  Each MD profile
is generated by a set of between 500 and 1300 individual ion
trajectories. 
Also shown
are the profiles obtained using the modified UT-MARLOWE BCA
code described in Sec.\ III. 
 Obviously, the MD calculation results are in
very good agreement with 
the experimental data, and with the BCA results.
This demonstrates that our electronic stopping power model 
provides a good physically-based damping mechanism for MD
simulations of ion implantation.

\section{Conclusion}
We have developed a phenomenological electronic stopping power
model for the physics of ion implantation. It has been
 implemented into MD and BCA Monte Carlo simulations.
SIMS data have been used to verify this model in the MD
and BCA Monte Carlo platforms. 
This model has only one free parameter, namely, the one
electron radius of unbound electrons in the medium. 
We
have fine tuned this parameter to obtain excellent results
of dopant profiles compared with SIMS data
in both MD and BCA Monte Carlo simulations.
We emphasize  that
this model with a single parameter can equally successfully
model both boron and arsenic implants into silicon
over a wide range of energies and in different
channeling
and off-axis directions of incidence. 
This
versatility indicates wide applicability of
the model in studies of other physical
processes involving electronic stopping.
As a more stringent test of the model, it should 
also be applied to
implantation of species other than boron and arsenic.
Using arsenic implantation as an example,
we have also addressed the issue of how significant
electronic stopping
is for heavy ions in a low energy regime. For instance, to
achieve a good quantitative understanding,
we still have to
take into account the physics of electronic stopping 
for arsenic
implants at 5keV.

As discussed above, it is important to incorporate ion-electron
couplings into MD simulations  in both the high energy radiation
damage regime and the low energy electron-phonon interaction
regime.
We have demonstrated that this model 
provides a crucial piece of physics in
MD simulations for modeling energetic collisions
in the electronic stopping power regime.
The agreement of the
simulated dopant profiles
with the SIMS data shows that the incorporation
of this physically-based damping term into MD simulations
is a phenomenologically reliable approach in the regime
concerned.
Under  way is an investigation of whether it can be used 
as a good phenomenological model for electron-phonon coupling
in the low energy regime.
This agreement also
suggests that MD can be used to generate dopant profiles
for testing against the low energy BCA results when 
experimental data is not available. Furthermore, MD simulations 
incorporating this physically-based damping mechanism
can provide 
valuable insight into how to modify the binary collision
approximation. This will enable 
 the validity of the Monte Carlo simulation to be extended
further into 
the  lower energy regime, while not destroying
computational efficiency required in realistic
simulation environments.

\section{Acknowledgment}
We thank Al Tasch for useful discussions and for providing us
with SIMS data, which facilitate  validation of the
model. This work is performed under the auspices of
the U.S. Department of Energy.

\protect\pagebreak
\section{Figures}

%\end{multicols}

\end{document}